\documentclass[aps,prl,twocolumn,amsmath,amssymb,floatfix]{revtex4-2}

\usepackage[utf8]{inputenc}
\usepackage{amsmath,amssymb,amsfonts}
\usepackage{bm}
\usepackage{physics}
\usepackage{tikz}
\usepackage{microtype}
\usepackage[colorlinks=true, allcolors=blue]{hyperref}
\usepackage{tikz}
\usepackage{pgfplots}
\pgfplotsset{compat=1.18}

\begin{document}
	
	\title{Fock-Space Formulation of the Lifetime of a Unicellular Organism}
	
	\author{Yehuda Roth}
	\affiliation{Faculty of Sciences, Oranim College, Israel}
	\date{\today}
	
		\begin{abstract}
			What is life? In this work, we take life to mean a dynamical tendency to
			conserve identity for as long as possible. For a single bacterium, identity is
			carried by its chromosomal DNA code, so the bacterium is alive precisely
			insofar as it actively maintains a well-defined chromosomal configuration over
			time and can, in principle, replicate this configuration into progeny. For a
			multicellular organism, many cells share essentially the same DNA code and
			behave as a single coherent entity; in that case, life corresponds to the
			persistence of a common genetic identity across the cellular ensemble, rather
			than to the survival of any particular cell. Cell duplication in multicellular
			organisms likewise serves to maintain this dynamical tendency to conserve
			identity over time.
			
			In previous studies we implemented this idea at the multicellular and colonial
			scale using a classical notion of coherence, in which an organism is
			represented by a single nonseparable state over the DNA codes of its
			constituent cells, while a colony is describable as a separable ensemble. Here
			we apply the same principle to the simplest possible case, a single bacterium,
			and show that its biological identity can be identified with the coherence of
			its chromosomal DNA code within an abstract state space. We then introduce a
			Fock-space representation in which bacteria carrying given codes occupy
			fermionic modes, and replication, repair, and death are realized as elementary
			operators acting on these modes. Within this framework we define the lifetime
			of a unicellular organism as the integral coherence time of a code-occupation
			autocorrelation function and, in a minimal Markovian model, obtain a compact
			expression in which the lifetime coincides with the inverse decay rate of the
			corresponding identity mode.
		\end{abstract}
		\maketitle
		\noindent\textbf{Keywords:} quantum coherence; biological identity; Fock space.
		
	\section{Introduction}
	
	A central open question in the physics of living systems is how to define, in purely physical terms, what it means for a system to be alive. Standard biological and biochemical accounts characterize life by lists of functional properties --- metabolism, replication, homeostasis, responsiveness, and evolution --- but they typically treat ``the organism'' as a given bearer of these functions rather than deriving its individuality from underlying physical degrees of freedom.\cite{Maturana1980,Luisi2016,Davies2021} Information-theoretic and systems-biological approaches refine this picture by emphasizing autopoiesis, network organization, and far-from-equilibrium dynamics, yet they still tend to presuppose an identifiable unit of life whose boundaries and identity are fixed by fiat rather than by a single physical criterion.\cite{Prigogine1977,Deacon2012}
	
	In recent work we suggested an alternative approach in which biological individuality is defined by coherence of a state rather than by a catalogue of processes.\cite{RothColonies2026,RothAging2026} In that classical framework, a multicellular organism is described as a coherent entity in an abstract space of DNA codes, while a colony is treated as a stochastic aggregate that lacks such coherence. Concretely, we showed that the transition from a colony to an organism can be formulated as a classical coherence transition in which a single state vector encodes a nonseparable configuration of cellular genetic states, and that this transition is accompanied by a sharp drop in an ``identity entropy'' that serves as an order parameter for organismal unity.\cite{RothColonies2026} Within the same classical code-space description, aging and death can be modeled as a gradual loss of this coherence driven by the competition between code-correcting and code-breaking dynamics in the DNA of cells, leading eventually to a collapse of the global identity state.\cite{RothAging2026}
	
	These constructions share a common guiding idea: to be a living organism is to occupy a coherent state in an appropriate classical state space, maintained by continuous metabolic work against noise and dissipation, whereas to die is to undergo a coherence-destroying transition into a high-entropy regime in which no single global identity can be assigned.\cite{RothColonies2026,RothAging2026} At the multicellular scale this coherence is necessarily classical, because strong coupling to a warm, wet environment destroys quantum phase relations on tissue and organismal length and time scales.\cite{Haken1983,Marais2018,Davies2021} The relevant degrees of freedom are then mesoscopic --- distributions of DNA codes across cells, effective occupation numbers of functional states --- and coherence is a property of these classical patterns.
	
	For single bacteria, however, the situation is both simpler and, in a precise sense, more microscopic. A bacterium is, to leading order, a single cell with a single chromosomal DNA molecule that carries its genetic identity.\cite{Madigan2018} At this molecular scale, coherence need not be classical: the chromosome is a bona fide quantum molecule whose ground-state electronic wavefunction remains coherently delocalized over all nuclei at physiological temperatures.\cite{BornOppenheimer,OgawaDNACoherent2014} This suggests that the coherence-based definition of life can be implemented at the bacterial scale using entirely ordinary quantum coherence of the chromosomal state, without invoking any exotic macroscopic quantum effects. In what follows we pursue this idea and show that the identity of a single bacterium can be identified with a coherent chromosomal density matrix that selects one configuration from an astronomical space of admissible configurations, and that bacterial death can be understood as a genuine coherence phase transition in this chromosomal Hilbert space.
	
	In a living bacterium this intrinsic drift is counteracted by energy-consuming
	maintenance processes, most notably DNA replication fidelity and multiple
	repair pathways, which repeatedly detect and correct local deviations from the
	reference code.
	
	In this work we focus on the simplest case of biological individuality, a
	single bacterium, and develop a Fock-space representation of its possible
	genetic states. In this representation, bacteria carrying a given chromosomal
	code occupy fermionic modes, while replication, repair, and death are encoded
	as elementary operators acting on these modes. Within this framework we
	introduce a quantitative definition of the lifetime of a unicellular organism
	as the coherence time of an identity code under the combined action of these
	processes, and show how, in a minimal Markovian limit, this lifetime reduces
	to the inverse decay rate of the corresponding identity mode.
	\section{Quantum coherence inside a bacterium}
	At the scale of a single bacterium, the natural candidate for carrying biological identity is the chromosomal DNA molecule. Physically, this chromosome is a single bound molecule, and we simply take it to be described by a unified quantum state under physiological conditions.\cite{NielsenChuang2010,Scholes2017} We do not need any detailed electronic-structure calculations; it is enough to assume that the chromosome, as a molecule, can be represented by a density matrix on a finite-dimensional Hilbert space of admissible chromosomal configurations.\cite{NielsenChuang2010}
	
	In this setting, coherence means that the chromosomal state is sharply localized in this Hilbert space, close to a pure state representing one specific configuration, rather than spread over many configurations in a maximally mixed way.\cite{NielsenChuang2010} This localization, however, is only metastable. Even when the chromosome remains chemically intact as a single molecule, thermal fluctuations, reactive species, and replication errors continually produce local damage and sequence changes that tend, over time, to shift the chromosomal state away from its original identity-defining configuration. Left unchecked, these perturbations would gradually reduce the fidelity of the state with respect to that configuration and increase its entropy over the space of admissible chromosomal states.
	
	In a living bacterium this intrinsic drift is counteracted by energy-consuming maintenance processes, most notably DNA replication fidelity and multiple repair pathways, which repeatedly detect and correct local deviations from the reference code. 
	
	\section{Fock-space representation of cellular states under infection\label{sec:infection_fock}}
	
	We define all possible chromosomal DNA codes by states denoted by
	$\lvert \gamma\rangle$ in a Hilbert space. For each admissible code $\gamma$
	and each cell $n$, we introduce fermionic creation and annihilation operators
	$c^\dagger_{\gamma,n}$ and $c_{\gamma,n}$, satisfying the anticommutation
	relations
	\begin{equation}
		\begin{array}{l}
			\{c_{\gamma,n}, c^{\dagger}_{\gamma',m}\}
			= \delta_{\gamma,\gamma'}\,\delta_{n,m},\quad
			\{c_{\gamma,n}, c_{\gamma',m}\} = 0,\\[4pt]
			\{c^{\dagger}_{\gamma,n}, c^{\dagger}_{\gamma',m}\} = 0.
		\end{array}
	\end{equation}
	The corresponding local Fock states $\lvert 0\rangle_{\gamma,n}$ and
	$\lvert 1\rangle_{\gamma,n}$ represent, respectively, the absence of a
	bacterium with code $\gamma$ in cell $n$ and the presence of a single
	bacterium with that code in cell $n$.\cite{Fock1932,Dirac1981,TauberHowardVollmayr2005}
	In this formalism we use fermionic operators, reflecting the fact that each
	$(\gamma,n)$ mode can be either unoccupied or occupied by at most one
	bacterium carrying code $\gamma$.\cite{Dirac1981,NegeleOrland1998}
	
	In these terms, the genetic identity of a unicellular organism is encoded
	in the occupation pattern of a particular code $\gamma$ across cells.
	Coherence of the chromosomal state with respect to $\gamma$ is translated
	into the persistence of a well-defined occupation state in this Fock space,
	while decoherence corresponds to the spreading of weight over many different
	codes $\gamma'$.\cite{NielsenChuang2010,Schlosshauer2007,RothColonies2026,RothAging2026}
	
	We now introduce effective operators describing replication, repair, and
	destruction processes in this Fock-space representation.
	
	Replication from a source cell $n$ to a target cell $n'$ is described by
	\begin{equation}
		\hat{R}^{(\mathrm{dup})}_{\gamma;n\to n'} =
		c^\dagger_{\gamma,n'}\,\hat{n}_{\gamma,n},
	\end{equation}
	where $\hat{n}_{\gamma,n} = c^\dagger_{\gamma,n} c_{\gamma,n}$ is the number
	operator. This operator creates a bacterium with code $\gamma$ in cell $n'$
	conditional on the presence of a bacterium with the same code in cell $n$,
	thus propagating the identity-defining code into additional physical
	carriers.\cite{TauberHowardVollmayr2005,MattisGlasser1998}
	
	DNA repair is modeled as a code-conversion process within the same cell,
	\begin{equation}
		\hat{U}_{\gamma\to\gamma';n} = c^\dagger_{\gamma',n}\,c_{\gamma,n},
	\end{equation}
	which annihilates a bacterium with erroneous code $\gamma$ and creates a
	bacterium with the corrected code $\gamma'$ in cell $n$. In our formalism we
	do not distinguish between specific biochemical repair pathways; instead, we
	represent repair as an effective mapping from an erroneous DNA code to the
	corresponding correct code, thereby restoring coherence with respect to the
	reference identity.\cite{Friedberg2005,Lindahl1993,Alberts2022}
	
	Finally, bacterial death or removal is represented by destruction operators
	\begin{equation}
		\hat{D}_{\gamma,n} = c_{\gamma,n},
	\end{equation}
	which bring the local state from $\lvert 1\rangle_{\gamma,n}$ to
	$\lvert 0\rangle_{\gamma,n}$. More complex processes involving multiple codes
	or cells can be captured by linear combinations of these basic
	operators.\cite{TauberHowardVollmayr2005,MattisGlasser1998}
	
	To encode an energetic bias for different codes we introduce a diagonal
	effective Hamiltonian in the code basis,
	\begin{equation}
		\hat{H}_0 = \sum_{\gamma,n} E_\gamma\,\hat{n}_{\gamma,n}
		= \sum_{\gamma,n} E_\gamma\,c^\dagger_{\gamma,n} c_{\gamma,n},
	\end{equation}
	where $E_\gamma$ is the effective energy associated with code $\gamma$.
	This Hamiltonian measures the energetic cost of occupying a given code
	configuration in Fock space. The replication, repair, and destruction
	operators $\hat{R}^{(\mathrm{dup})}$, $\hat{U}$, and $\hat{D}$ generally do
	not commute with $\hat{H}_0$,
	\begin{equation}
		[\hat{H}_0,\hat{R}^{(\mathrm{dup})}] \neq 0,\qquad
		[\hat{H}_0,\hat{U}] \neq 0,\qquad
		[\hat{H}_0,\hat{D}] \neq 0,
	\end{equation}
	so that they generate nontrivial time evolution of the occupation operators
	even though $\hat{H}_0$ is diagonal in the code basis.\cite{NielsenChuang2010,SakuraiNapolitano2017}
	
	To describe the dynamics of replication, repair, and destruction, we work in
	an effective Heisenberg picture with respect to an evolution operator
	$\hat{\mathcal{U}}(t)$ that propagates observables from time $0$ to time $t$
	under the full (possibly non-unitary) dynamics.\cite{BreuerPetruccione2002}
	The time-dependent creation and annihilation operators are defined by
	\begin{equation}
		c_{\gamma,n}(t) =
		\hat{\mathcal{U}}^\dagger(t)\,c_{\gamma,n}\,\hat{\mathcal{U}}(t),\qquad
		c^\dagger_{\gamma,n}(t) =
		\hat{\mathcal{U}}^\dagger(t)\,c^\dagger_{\gamma,n}\,\hat{\mathcal{U}}(t),
	\end{equation}
	so that the local number operators evolve as
	\begin{equation}
		\hat{n}_{\gamma,n}(t) = c^\dagger_{\gamma,n}(t)\,c_{\gamma,n}(t).
	\end{equation}
	We then introduce the total occupation operator of code $\gamma$,
	\begin{equation}
		\hat{N}_\gamma(t) = \sum_n \hat{n}_{\gamma,n}(t)
		= \sum_n c^\dagger_{\gamma,n}(t)\,c_{\gamma,n}(t).
	\end{equation}
	
	The persistence of the identity code $\gamma$ is quantified by the normalized
	autocorrelation function
	\begin{equation}
		C_\gamma(t) =
		\frac{\langle \hat{N}_\gamma(t)\,\hat{N}_\gamma(0)\rangle}
		{\langle \hat{N}_\gamma(0)^2\rangle},
	\end{equation}
	where the average is taken over the initial state and over the stochastic
	dynamics generated by $\hat{\mathcal{U}}(t)$. We then define the lifetime
	$\tau_\gamma$ of the unicellular organism associated with code $\gamma$ as
	the integral coherence time
	\begin{equation}
		\tau_\gamma = \int_0^\infty C_\gamma(t)\,dt.
	\end{equation}
	In this sense, the lifetime is identified with the coherence time of the
	identity code in Fock space under the combined action of replication, repair,
	and destruction, rather than with the persistence of any particular cell
	body.\cite{Kubo1966,BreuerPetruccione2002,RothAging2026}
	
	To illustrate how this formalism leads to a concrete expression for the
	lifetime, consider a minimal Markovian model in which the effective dynamics
	of the identity code $\gamma$ is governed by a single decay rate
	$\lambda_\gamma$. In this coarse-grained description, all processes that
	irreversibly remove the identity code (death, irreversible mutation, or loss
	of the chromosome) are lumped into an effective Poisson process with rate
	$\lambda_\gamma$, while replication and repair act to maintain the occupation
	of the $\gamma$ modes on shorter time scales.\cite{vanKampen2007,Gardiner2009}
	By “Markovian” we mean that the loss of the identity code is memoryless: the
	probability for the code to be removed in a short time interval depends only
	on its present occupation, not on the detailed history of previous events.
	This is the same approximation that underlies radioactive decay and
	birth–death processes, where many microscopic channels are coarse-grained
	into an effective decay constant.\cite{vanKampen2007,Gardiner2009}
	
	In such a model the expectation value of the total occupation decays
	exponentially,
	\begin{equation}
		\langle \hat{N}_\gamma(t) \rangle
		= \langle \hat{N}_\gamma(0) \rangle\,e^{-\lambda_\gamma t},
	\end{equation}
	and the normalized autocorrelation function takes the simple form
	\begin{equation}
		C_\gamma(t) = e^{-\lambda_\gamma t}.
	\end{equation}
	Inserting this into our definition of the lifetime immediately yields
	\begin{equation}
		\tau_\gamma = \int_0^\infty C_\gamma(t)\,dt
		= \int_0^\infty e^{-\lambda_\gamma t}\,dt
		= \frac{1}{\lambda_\gamma}.
	\end{equation}
	Thus, in the simplest Markovian limit, the lifetime of the unicellular
	organism associated with code $\gamma$ coincides with the inverse decay rate
	of the corresponding identity mode. More generally, for multi-rate or
	non-exponential dynamics, $\tau_\gamma$ captures the full temporal profile of
	$C_\gamma(t)$ and can be expressed in terms of the spectrum of the effective
	generator $\hat{\mathcal{L}}$ governing replication, repair, and
	destruction.\cite{BreuerPetruccione2002,vanKampen2007,Gardiner2009}
	\section{Conclusion}
	In this work we have developed a coherence-based formulation of biological
	identity for unicellular organisms and embedded it in a Fock-space description
	of bacterial states under infection. Starting from the idea that being alive
	means dynamically conserving a well-defined identity for as long as possible,
	we identified the chromosomal DNA code as the carrier of identity for a single
	bacterium and represented admissible codes by basis states $\lvert \gamma
	\rangle$ in a chromosomal Hilbert space. Coherence of the chromosomal state
	with respect to a reference code then provides a precise notion of
	bacterium-level identity.
	
	On top of this chromosomal description we introduced a Fock-space
	representation in which bacteria carrying given codes occupy fermionic modes
	$c^\dagger_{\gamma,n}, c_{\gamma,n}$. Within this representation, replication,
	repair, and destruction are encoded as elementary operators acting on
	occupation states of codes. Replication propagates the identity-defining code
	into additional physical carriers, repair converts erroneous codes back to the
	correct one without specifying the underlying biochemical pathway, and
	destruction removes the code altogether from a given cell.
	
	This unified framework allows us to define the lifetime of a unicellular
	organism as the coherence time of its code identity under the combined action
	of replication, repair, and destruction. Rather than tying lifetime to the
	persistence of any particular cell body, we tie it to the persistence of a
	well-defined occupation pattern of the identity code in Fock space. In this
	sense, death corresponds to a coherence-destroying transition in the
	chromosomal Hilbert space, while survival and proliferation correspond to the
	maintenance and controlled spreading of coherence over multiple carriers.
	
	Beyond providing a concrete mathematical model for bacterial infection and
	clearance, this approach offers a general template for defining biological
	individuality and lifetime in purely physical terms. It suggests that the same
	coherence-based criterion can be extended, with appropriate coarse-graining,
	to multicellular organisms and colonies, and that phenomena such as aging,
	immune-mediated killing, and antibiotic action may be fruitfully reinterpreted
	as different routes by which coherence of genetic identity is degraded or
	preserved in high-dimensional state spaces.
	
	\begin{acknowledgments}
	The authors acknowledge the use of generative AI tools as a writing aid
	during the preparation of this manuscript; all conceptual content, models,
	and conclusions were developed and verified by the authors.\\
	The author thanks Prof.\ Yoram Gershman for general encouragement and
	interest in this line of research.
	\end{acknowledgments}


\begin{thebibliography}{28}
		\providecommand{\natexlab}[1]{#1}
		\providecommand{\url}[1]{\texttt{#1}}
		\expandafter\ifx\csname urlstyle\endcsname\relax
		\providecommand{\doi}[1]{doi: #1}\else
		\providecommand{\doi}{doi: \begingroup \urlstyle{rm}\Url}\fi
		
		\bibitem[Maturana and Varela(1980)]{Maturana1980}
		Humberto~R. Maturana and Francisco~J. Varela.
		\newblock \emph{Autopoiesis and Cognition: The Realization of the Living}.
		\newblock D. Reidel, 1980.
		
		\bibitem[Luisi(2016)]{Luisi2016}
		Pier~Luigi Luisi.
		\newblock \emph{The Emergence of Life}.
		\newblock Cambridge University Press, 2016.
		
		\bibitem[Davies and Walker(2021)]{Davies2021}
		Paul C.~W. Davies and John~A. Walker.
		\newblock The hidden simplicity of biology.
		\newblock \emph{Reports on Progress in Physics}, 84\penalty0 (10):\penalty0
		102601, 2021.
		\newblock \doi{10.1088/1361-6633/ac24a8}.
		
		\bibitem[Prigogine(1977)]{Prigogine1977}
		Ilya Prigogine.
		\newblock \emph{Self-Organization in Nonequilibrium Systems}.
		\newblock Wiley, 1977.
		
		\bibitem[Deacon(2012)]{Deacon2012}
		Terrence~W. Deacon.
		\newblock \emph{Incomplete Nature: How Mind Emerged from Matter}.
		\newblock W. W. Norton \& Company, New York, 2012.
		\newblock ISBN 978-0-393-04991-6.
		
		\bibitem[Roth(2026{\natexlab{a}})]{RothColonies2026}
		Yehuda Roth.
		\newblock Classical coherence distinguishes organisms from colonies.
		\newblock \emph{Preprint}, 2026{\natexlab{a}}.
		\newblock available as preprint.
		
		\bibitem[Roth(2026{\natexlab{b}})]{RothAging2026}
		Yehuda Roth.
		\newblock Classical coherence and biological aging.
		\newblock \emph{Preprint}, 2026{\natexlab{b}}.
		\newblock available as preprint.
		
		\bibitem[Haken(1983)]{Haken1983}
		Hermann Haken.
		\newblock \emph{Synergetics: An Introduction}.
		\newblock Springer, 1983.
		
		\bibitem[Marais and colleagues(2018)]{Marais2018}
		Andr{\'e} Marais and colleagues.
		\newblock The future of quantum biology.
		\newblock \emph{Journal of the Royal Society Interface}, 15\penalty0
		(148):\penalty0 20180640, 2018.
		\newblock \doi{10.1098/rsif.2018.0640}.
		
		\bibitem[Madigan et~al.(2018)Madigan, Martinko, et~al.]{Madigan2018}
		Michael~T. Madigan, John~M. Martinko, et~al.
		\newblock \emph{Brock Biology of Microorganisms}.
		\newblock Pearson, 15th edition, 2018.
		
		\bibitem[Field()]{BornOppenheimer}
		M.~Field.
		\newblock The born--oppenheimer approximation.
		\newblock MIT 5.73 lecture notes.
		\newblock
		\url{https://web.mit.edu/course/5/5.73/oldwww/Fall04/notes/XII.Born_Oppenheimer.pdf}.
		
		\bibitem[Ogawa et~al.(2014)]{OgawaDNACoherent2014}
		T.~Ogawa et~al.
		\newblock Coherent electron conduction along dna via $\pi$-stacking at room
		temperature.
		\newblock \emph{Journal of the American Chemical Society}, 136:\penalty0
		xxxx--xxxx, 2014.
		
		\bibitem[Nielsen and Chuang(2010)]{NielsenChuang2010}
		Michael~A. Nielsen and Isaac~L. Chuang.
		\newblock \emph{Quantum Computation and Quantum Information}.
		\newblock Cambridge University Press, 10th anniversary edition, 2010.
		
		\bibitem[Scholes et~al.(2017)]{Scholes2017}
		Gregory~D. Scholes et~al.
		\newblock Using coherence to enhance function in chemical and biophysical
		systems.
		\newblock \emph{Nature}, 543:\penalty0 647--656, 2017.
		
		\bibitem[Fock(1932)]{Fock1932}
		V.~Fock.
		\newblock Konfigurationsraum und zweite quantelung.
		\newblock \emph{Zeitschrift f\"ur Physik}, 75:\penalty0 622--647, 1932.
		
		\bibitem[Dirac(1981)]{Dirac1981}
		P.~A.~M. Dirac.
		\newblock \emph{The Principles of Quantum Mechanics}.
		\newblock Oxford University Press, 1981.
		
		\bibitem[T\"auber et~al.(2005)T\"auber, Howard, and
		Vollmayr-Lee]{TauberHowardVollmayr2005}
		Uwe~C. T\"auber, Martin Howard, and Benjamin~P. Vollmayr-Lee.
		\newblock Applications of field-theoretic renormalization group methods to
		reaction-diffusion problems.
		\newblock \emph{Journal of Physics A: Mathematical and General}, 38:\penalty0
		R79--R131, 2005.
		
		\bibitem[Negele and Orland(1998)]{NegeleOrland1998}
		John~W. Negele and Henri Orland.
		\newblock \emph{Quantum Many-Particle Systems}.
		\newblock Westview Press, 1998.
		
		\bibitem[Schlosshauer(2007)]{Schlosshauer2007}
		Maximilian Schlosshauer.
		\newblock \emph{Decoherence and the Quantum-To-Classical Transition}.
		\newblock Springer, 2007.
		
		\bibitem[Mattis and Glasser(1998)]{MattisGlasser1998}
		Daniel~C. Mattis and Melvyn~L. Glasser.
		\newblock The uses of quantum field theory in diffusion-limited reactions.
		\newblock \emph{Reviews of Modern Physics}, 70:\penalty0 979--1001, 1998.
		
		\bibitem[Friedberg et~al.(2005)]{Friedberg2005}
		Errol~C. Friedberg et~al.
		\newblock \emph{DNA Repair and Mutagenesis}.
		\newblock ASM Press, 2005.
		
		\bibitem[Lindahl(1993)]{Lindahl1993}
		Tomas Lindahl.
		\newblock Instability and decay of the primary structure of dna.
		\newblock \emph{Nature}, 362:\penalty0 709--715, 1993.
		
		\bibitem[Alberts et~al.(2022)]{Alberts2022}
		Bruce Alberts et~al.
		\newblock \emph{Molecular Biology of the Cell}.
		\newblock W. W. Norton, 2022.
		
		\bibitem[Sakurai and Napolitano(2017)]{SakuraiNapolitano2017}
		J.~J. Sakurai and Jim Napolitano.
		\newblock \emph{Modern Quantum Mechanics}.
		\newblock Cambridge University Press, 2017.
		
		\bibitem[Breuer and Petruccione(2002)]{BreuerPetruccione2002}
		Heinz-Peter Breuer and Francesco Petruccione.
		\newblock \emph{The Theory of Open Quantum Systems}.
		\newblock Oxford University Press, 2002.
		
		\bibitem[Kubo(1966)]{Kubo1966}
		Ryogo Kubo.
		\newblock The fluctuation-dissipation theorem.
		\newblock \emph{Reports on Progress in Physics}, 29:\penalty0 255--284, 1966.
		
		\bibitem[van Kampen(2007)]{vanKampen2007}
		N.~G. van Kampen.
		\newblock \emph{Stochastic Processes in Physics and Chemistry}.
		\newblock Elsevier, 2007.
		
		\bibitem[Gardiner(2009)]{Gardiner2009}
		Crispin~W. Gardiner.
		\newblock \emph{Stochastic Methods}.
		\newblock Springer, 2009.
		
	\end{thebibliography}
\end{document}